\documentstyle[12pt]{article}
\begin{document}
%%%%%%%%%%%%%%%%%%%%%%%%%%%%%%%%%%%%%%%%%%%%%%%%%%%%%%%%%%%%%%%%%%%%%%%%%%%%%

\large

\begin{center}
{\Large \bf
{ NEW MECHANISM OF $CP$-VIOLATION } \\
{ IN THE MODELS OF ELECTROWEAK BARYOGENESIS }
}\\
\vskip 0.3cm
{ Andro BARNAVELI and Merab GOGBERASHVILI } \\
\vskip 0.3cm
{\it
 {Institute of Physics, Georgian Acad. Sci.}\\
 {Tamarashvili str. 6, Tbilisi 380077, Republic of Georgia.}\\
 {(E-mail: bart@physics.iberiapac.ge ; gogber@physics.iberiapac.ge ).}
} \\
\vskip 0.5cm

{\Large \bf Abstract}\\
\vskip 0.3cm
\quotation

\normalsize

The new source of $CP$-violation in the frames of "charge transport"
mechanism of electroweak baryogenesis is investigated. The consideration is
based on the assumption that $C$ and $CP$ need not be violated at the same
place, and on the fact that in the case of nonpolarized flux of
fermions falling nonperpendicularily on the phase-separating domain wall the
transmitted flux will be polarized  (i.e. $P$ and $CP$ are violated).
Sphaleron processes can convert the ($P$ and $CP$)-asymmetric state into
($C$ and $CP$)-asymmetric one. We argue that the value of $CP$-violation in
such mechanism can explain the observed value of Baryon Asymmetry of the
Universe even in the frames of the Minimal Standard Model.

\endquotation
\end{center}
\vskip 1.0cm

\large
%%%%%%%%%%%%%%%%%%%%%%%%%%%%%%%%%%%%%%%%%%%%%%%%%%%%%%%%%%%%%%%%%%%%%%%%%%%%

\def\theequation{\arabic{section}.\arabic{equation}}

\section{Introduction}
\setcounter{equation}{0}

{}~~~~~Almost all models describing generation of Baryon Asymmetry of the
Universe (BAU) on the electroweak scale use different extensions of the
Minimal Standard Model (MSM), since the value of $C$- and $CP$-violation
(which is necessary for baryogenesis) following from Kobayashi-Maskawa
matrix is too small to explain the observed value of BAU \cite{CKN93,T92}

\begin{equation}
\delta = \frac{n_B}{s} \sim 10^{-11} \div 10^{-8}
\label{1.1}
\end{equation}
(where $n_B$ is the net baryon density and $s$ is the entropy density).
The only exception is the model elaborated in works \cite{S91,FS93}, which
tries to explain BAU in the frames of MSM and uses temperature effects to
yield the sufficient $CP$-violation. However it causes a lot of discussions
\cite{GHOP93} and the question remains open.

In this paper we consider the other possible source of $C$- and $CP$-
violation which arises during the first order electroweak phase transitions
even in the frames of MSM.

We are basing on the following moments:
\begin{enumerate}
\item
{ {\it For the BAU generation $C$- and $CP$-violation need not occur at the
same place}. This assumption is analogous to that given in the so-called
"charge transport" mechanism \cite{CKN91,CKN92}, where it was considered
that $B$- and $CP$-violation take place during the first order phase
transitions in the different areas: $CP$-violation --- exactly on the
phase-separating domain wall, while $B$-violation --- in the symmetric
phase.
}
\item
{ {\it Interaction of particles with domain wall can violate $P$- and
$CP$-symmetry}. Indeed,  as we will see, if one considers the nonpolarized
flux of fermions (i.e. the equal numbers of positive- and negative-helicity
fermions) falling not perpendicularly upon a phase-separating domain wall,
then the transmitted flux will be polarized i.e. ($P$ and $CP$)-asymmetric.
Note, that the $P$-violating features of domain walls are well
known (see \cite{CH} and references therein). Here we would like to mention
also, that the fenomenon of fermion flux polarization by the phase boundary
is in some sense analogous to the "Mallus effect" in optics \cite{S} when
the nondirect sun rays are polarized by the atmosphere.  }

\item
{ {\it Any $P$-asymmetry inside the  new-phase bubble is eliminated} due to
multiple interactions of massive particles and with account of spherical
symmetry of the problem.  }
\item
{ {\it The  anomalous $B$-violating sphaleron processes in the symmetric
phase can convert the axial asymmetry into baryon asymmetry} \cite{CKN92}.
Thus, effectively, ($P$ and $CP$)-asymmetry (created on the domain wall) is
converted (in the symmetric phase) into ($C$ and $CP$)-asymmetry.
}
\end{enumerate}

Here we would like to note that some misunderstanding can take place, since
the bubble solutions are spherically symmetric and thus, globally, no
$P$-violation must occur.  However nucleation of bubbles creates the
distinguished frame, for example, the rest frame of the bubble wall and
violates Lorentz-invariance. Namely in this frame parity is violated and
such "local" $P$-violation plays the crucial role in our mechanism. Besides,
due to processes inside and outside the bubbles this $P$-asymmetry is
eliminated or converted into ($C$ and $CP$)-asymmetry. Thus, globally, no
$P$-asymmetry remains and this problem must not arise.

In the next section we describe shortly the mechanism of baryogenesis.

In section 3 the parity violation in the fermion -- wall interactions is
considered.

In section 4 the numerical estimation of BAU, generated through the
described mechanism, is given.

%%%%%%%%%%%%%%%%%%%%%%%%%%%%%%%%%%%%%%%%%%%%%%%%%%%%%%%%%%%%%%%%%%%%%%%%%

\section{The mechanism of baryogenesis.}
\setcounter{equation}{0}

{}~~~~~The mechanism of baryogenesis, described above, works at the
first order phase transitions. Such phase transitions proceed via nucleation
and expansion of the new (broken) phase bubbles within the old (symmetric)
one \cite{CKN93,HKLLM92,T91}. These bubbles grow until the universe is
completely converted to the broken phase.

During their motion through the plasma the bubble walls interact with the
particles. This interactions violate $P$ and $CP$. Namely, for the fermions
falling nonperpendicularly upon the wall the latter changes the polarization
of initial flux.  With the natural assumption that the equal number of
right- and left-handed fermions from the old phase are colliding with the
phase boundary this will cause the polarized flux of particles penetrating
into the broken phase (i.e. more fermions of, say, negative helicity enter
the new phase through the bubble wall).

Since the mean radius of bubbles (to the end of the phase transition) $R\sim
10^{12}/T$ \cite{T91} is by many orders greater than the particle mean free
path $l\sim (4\div 10)/T$ \cite{CKN93}, the fermions penetrated into the
new phase bubble can reach its walls again only after collisions with other
particles. This interactions of massive particles (with account of spherical
symmetry of the problem) eliminate the helicity-asymmetry (i.e.
$P$-asymmetry) inside the bubbles and one can assume that equal numbers of
positive- and negative-helicity fermions (i.e. nonpolarized flux) from the
new phase do reach the bubble walls.

Now let us see, what does happen with the nonpolarized flux of fermions
falling upon the phase boundary from the new phase, i.e. from inside the
bubble (of course, due to the wall motion, this flux is smaller than that
from the unbroken phase).  Due to polarization-changing features of domain
walls the polarized chiral (or "axial") flux of fermions coming out of the
bubble into the symmetric phase will occur. The amount of right-handed
fermions (which are isosinglets) and anti-fermions (which are isodoublets)
entering the symmetric phase will be greater than the amount of left-handed
fermions (which are isodoublets) and anti-fermions (which are isosinglets)
entering it.  Note that for the massless particles in the symmetric phase
the definitions of "helicity" and "chirality" are equivalent; the positive
helicity corresponds to the right chirality and negative helicity --- to the
left chirality. Thus one can say that the new phase bubble emits into an
unbroken phase a ($P$ and $CP$)-asymmetric distribution of particles.

The particle distribution functions in front of the wall approach local
thermal and chemical equilibrium, characterized by a temperature and by
chemical potentials for the various conserved and approximately conserved
charges. Because isosinglet and isodoublet particles carry different values
of weak hypercharge, which is a conserved quantum number in the unbroken
phase, the nonzero "axial" flux from the wall will lead to a nonzero density
of fermionic hypercharge $Y$ in the region preceding the wall \cite{CKN92}.
Using of this quantum number is more convenient, since the "axial charge" is
not a good quantum number, as it is violated by Higgs scattering
\cite{CKN93}. Note, that with account of charge-screening effects it will be
more correct to consider not the fermionic hypercharge $Y$, but rather the
linear combination $B' = B - xY$, where $B$ is the baryon number and $x$ is
chosen so that $B'$ is not screened by gauge forces (calculations show that
$x\sim 1/5$ \cite{CKN93,CKN92}). However this does not change the results
significantly.

As it was shown \cite{CKN93,CKN91,CKN92,AB94}, in the presence of nonzero
hypercharge the free energy of the unbroken phase is minimized for nonzero
baryon number. Anomalous baryon violating electroweak processes will be
biased favoring the production of baryon number over anti-baryon number.
Therefore the ($P$ and $CP$)-asymmetric state will be converted into ($C$
and $CP$)-asymmetric state.

Then this baryon asymmetry, created in the symmetric phase, will eventually
pass into the expanding bubbles of the broken phase, where baryons are
stable, and is conserved up to day.

%%%%%%%%%%%%%%%%%%%%%%%%%%%%%%%%%%%%%%%%%%%%%%%%%%%%%%%%%%%%%%%%%%%%%%%%

\section{Interaction of fermions with the phase boundary.}
\setcounter{equation}{0}

{}~~~~~We have noted above, that interactions of fermions with the phase
boundary violate ($P$ and $CP$)-invariance. This is exhibited in the
difference between polarizations of the initial and the transmitted fluxes.
Let us discuss this subject in more details.

Fermions are interacting with the scalar field, forming the bubble, through
the Yukawa coupling. Since in the old phase the vacuum expectation value
(VEV) of the scalar field $<\varphi >=0$ and in the new phase $<\varphi
>\neq 0$, the mass of fermions is spatially varying. In fact, one has to
deal with the Dirac equation with spatially varying mass term.

\begin{equation}
\left(\gamma_\mu \partial^\mu - m(z)\right) \psi = 0 .
\label{3.1}
\end{equation}
To solve such equation we shall make some simplifications.

First of all, since the bubble radius is by many orders larger than the
particle mean free path, we can consider the bubble wall as the planar one
(e.g. in the x-y plane) and to assume, that it moves in the negative $z$
direction. The second, at the particular choice of parameters, in MSM, the
wall width can be $\delta_w \sim O(1/T)$ \cite{HKLLM92}, i.e. smaller, than
the particle mean free path. Then one can neglect $\delta_w$ and consider
the wall as infinitely thin. In this case one has:

\begin{equation}
m(z) = \left\{ { {0, \quad z<0} \atop {m, \quad z>0 .} } \right.
\label{3.2}
\end{equation}

In the rest frame of the bubble wall the problem is equivalent to the
quantum mechanical problem of fermion scattering from the rectangular
potential barrier \cite{F74}. We shall investigate the nonperpendicular
scattering. Note, that in all models of "charge transport" mechanism
\cite{CKN91,CKN92} and also in \cite{FS93} only normal scattering of
particles from the phase boundary was considered, while namely
nonperpendicular scattering effects give the nonzero contribution to the
$CP$-violation.

Let us introduce the new parameter

\begin{equation}
\eta \equiv \frac{E-m}{k} = \frac{k}{E+m} = \sqrt{\frac{E-m}{E+m}}
\label{3.3}
\end{equation}
(where $E$ is the fermion energy and $\vec k$ --- its momentum),  the
angle $\theta$ characterizing the direction of $\vec k$:

\begin{equation}
k_t \equiv k_x = k\sin\theta \quad ; \quad k_l\equiv k_z = k\cos\theta
\label{3.4}
\end{equation}
(where $k_l, k_t$ are the momenta perpendicular and parallel to the wall
respectively; due to cylindrical symmetry of the problem the direction of
$\vec k$ does not depend on the polar angle $\phi$  and one can consider
scattering in x-z plane), and denote the incident, reflected and transmitted
waves as $\psi$, $\psi '$ and $\psi ''$ respectively, while their wave
vectors as $\vec k$, $\vec k'$ and $\vec k''$ and the angles characterizing
their direction as $\theta$, $\theta '$ and $\theta ''$ respectively. At the
phase boundary $(z=0)$ the following condition must be satisfied

\begin{equation}
\psi + \psi ' = \psi ''.
\label{3.6}
\end{equation}
This implies that

\begin{eqnarray}
\theta ' = \pi - \theta \nonumber \\
k \sin \theta = k''\sin\theta '' .
\label{3.7}
\end{eqnarray}

Let us assume that the incoming flux of particles is nonpolarized i.e. equal
amounts of positive- and negative-helicity fermions are falling upon the
wall. Then in the standard basis

\begin{equation}
 \psi = \frac{1}{\sqrt{V(1+\eta^2)}} \cdot
 \left( \begin{array}{c}
 \left( \cos \frac{\theta}{2} -\sin \frac{\theta}{2} \right)  \\
 \left( \sin \frac{\theta}{2} +\cos \frac{\theta}{2} \right)  \\
 \eta \left( \cos \frac{\theta}{2} +\sin \frac{\theta}{2} \right)  \\
 \eta \left( \sin \frac{\theta}{2} -\cos \frac{\theta}{2} \right)
 \end{array} \right)
 \cdot e^{-i(Et - \vec k\vec r)} .
\label{3.8}
\end{equation}
Here $V$ is the volume of normalization and we use the signature
$(+,-,-,-)$.
The reflected and the transmitted waves can be written in the form

\begin{equation}
 \psi ' = \frac{1}{\sqrt{V(1+\eta^2)}} \cdot
 \left( \begin{array}{c}
 \left( -C\sin \frac{\theta}{2} +D\cos \frac{\theta}{2} \right)  \\
 \left( C\cos \frac{\theta}{2} +D\sin \frac{\theta}{2} \right)  \\
 \eta \left( C\sin \frac{\theta}{2} +D\cos \frac{\theta}{2} \right)  \\
 \eta \left( -C\cos \frac{\theta}{2} +D\sin \frac{\theta}{2} \right)
 \end{array} \right)
 \cdot e^{-i(Et - \vec k'\vec r)} ;
\label{3.9}
\end{equation}

\begin{equation}
 \psi '' = \frac{1}{\sqrt{V(1+\eta^2)}} \cdot
 \left( \begin{array}{c}
 \left( G\cos \frac{\theta ''}{2} -F\sin \frac{\theta ''}{2} \right)  \\
 \left( G\sin \frac{\theta ''}{2} +F\cos \frac{\theta ''}{2} \right)  \\
 \eta '' \left( G\cos \frac{\theta ''}{2} +F\sin \frac{\theta ''}{2} \right)
\\
 \eta ''\left( G\sin \frac{\theta ''}{2} -F\cos \frac{\theta ''}{2} \right)
 \end{array} \right)
 \cdot e^{-i(Et - \vec k''\vec r)}  ,
\label{3.10}
\end{equation}
respectively, where the coefficients $C$ and $G$ characterize the states
with positive helicity, while coefficients $D$ and $G$ --- states with
negative helicity and $\eta = \eta '$. Substituting equations
(\ref{3.8}),(\ref{3.9}),(\ref{3.10}) into boundary condition (\ref{3.6}) one
can yield \cite{F74}:

\begin{equation}
C = \frac{-\alpha + \beta}{\Delta} \quad ; \quad
D = -\frac{\alpha + \beta}{\Delta} \quad ;
\label{3.11}
\end{equation}

\begin{equation}
G = \frac{\rho + \sigma}{2\xi\lambda} \quad ; \quad
F = \frac{\rho - \sigma}{2\xi\lambda} \quad ;
\label{3.12}
\end{equation}
where

\begin{eqnarray}
\alpha = 2(\lambda^2+1)p(1+q^2) - 4\lambda q(1+p^2) , \nonumber \\
\beta = (\lambda^2-1)(1-p^2)(1+q^2) , \nonumber \\
\rho = \frac{1}{\Delta}\cdot 4\lambda (\lambda +1)(1-pq)(1-p^2) , \nonumber
\\
\sigma = \frac{1}{\Delta}\cdot 4\lambda (\lambda -1)(p+q)(1-p^2) ,
\nonumber \\
\Delta = (\lambda +1)^2(1-pq)^2 + (\lambda -1)^2(p+q)^2 , \nonumber \\
p \equiv \tan \frac{\theta}{2} \quad , q \equiv \tan \frac{\theta ''}{2}
\quad , \nonumber \\
\xi = \sqrt{\frac{1+\eta^2}{1+\eta ''^2}}\cdot
\frac{\cos\frac{\theta ''}{2}}{\cos\frac{\theta}{2}} ; \nonumber \\
\lambda = \frac{\eta ''}{\eta} .
\label{3.13}
\end{eqnarray}
It is easy to understand that polarization of reflected and transmitted
fluxes are

\begin{eqnarray}
h' = \frac{C^2-D^2}{C^2+D^2} = - \frac{2\alpha\beta}{\alpha^2 + \beta^2},
\nonumber \\
h'' = \frac{G^2-F^2}{G^2+F^2} = \frac{2\rho\sigma}{\rho^2 + \sigma^2} .
\label{3.14}
\end{eqnarray}

If one considers the flux of particles coming towards the phase boundary
from the unbroken phase, then $\eta = \eta ' =1, \;$ $\eta '' < 1$. In this
case, with account that $0 < \theta /2 <\pi /4$ i.e. $p <1$ and $q < 1$, one
yields: $\alpha = 0, \; \lambda <1, \; \rho >0, \; \sigma <0$. This  means
that $h' = 0$ and $h'' <0$, i.e. the reflected flux will be nonpolarized,
while the transmitted flux will have a negative polarization.

Now, if one considers the flux of particles falling on the wall from the new
phase, then $k = k' = \sqrt{E^2 - m^2}, \; k''=E, \; \eta = \eta ' <1, \;
\eta '' =1$. Then $\lambda >1, \; \sigma >1, \;$ and $h'' >0$. This means
that the positively polarized flux of fermions (i.e. more right-handed
fermions than those left-handed) will be emitted from the new phase bubble
into the old phase.

Note that if the initial flat wave falls on the wall perpendicularly, then
$\theta = \theta '' = 0 \; \Rightarrow \; p=q=0 \; \Rightarrow \; \alpha =
\sigma =0 \; \Rightarrow \; h' = h'' =0$ and the polarization of reflected
and transmitted fluxes will be equal to that of the initial flux.

Now let us calculate the coefficients of reflection and transmission through
the phase boundary. We can express the current density

\begin{equation}
j_\mu = ie\bar{\psi} \gamma_\mu\psi
\label{3.15}
\end{equation}
in terms of spinor components. Using (\ref{3.8}) -- (\ref{3.13}) one can
obtain that

\begin{eqnarray}
j_x = \frac{4e\eta}{V(1+\eta^2)}\cdot \sin\theta ; \nonumber \\
j_z = \frac{4e\eta}{V(1+\eta^2)}\cdot \cos\theta .
\label{3.16}
\end{eqnarray}

\begin{eqnarray}
j'_x = \frac{2e\eta}{V(1+\eta^2)}\cdot (C^2+D^2) \sin\theta ; \nonumber \\
j'_z = -\frac{2e\eta}{V(1+\eta^2)}\cdot (C^2+D^2) \cos\theta .
\label{3.17}
\end{eqnarray}

\begin{eqnarray}
j''_x = \frac{2e\eta ''}{V(1+\eta ''^2)}\cdot (G^2+F^2) \sin\theta '' ;
\nonumber \\
j''_z = \frac{2e\eta ''}{V(1+\eta ''^2)}\cdot (G^2+F^2) \cos\theta ''.
\label{3.18}
\end{eqnarray}
Let us define the "full coefficient of reflection" as the ratio of reflected
current density to the incident one:

\begin{equation}
R \equiv \frac{\sqrt{{j'_x}^2 + {j'_z}^2}}{\sqrt{j_x^2 + j_z^2}} =
\frac{C^2 + D^2}{2},
\label{3.19}
\end{equation}
and the "full coefficient of transmission" as the ratio of transmitted
current density to the incident one:

\begin{equation}
K \equiv \frac{\sqrt{{j''_x}^2 + {j''_z}^2}}{\sqrt{j_x^2 + j_z^2}} =
\frac{\eta ''(1+\eta^2)}{\eta (1+{\eta ''}^2)}\cdot\frac{(G^2 + F^2)}{2}.
\label{3.20}
\end{equation}

%%%%%%%%%%%%%%%%%%%%%%%%%%%%%%%%%%%%%%%%%%%%%%%%%%%%%%%%%%%%%%%%%%%%%%%%

\section{Computation of BAU.}
\setcounter{equation}{0}

{}~~~~~Now we can compute the net flux of axial number coming from the new
phase bubble into the symmetric phase.

Taking into account, that both --- the left-handed fermions which are
isodoublets and the right-handed fermions which are isosinglets, and also
their antiparticles -- left-handed antifermions which are isosinglets and
right-handed antifermions which are isodoublets, can be emitted into
unbroken phase (we denote the plasma-frame fluxes of these species
as $(f^d_L)$, $\; (f^s_R)$, $\; (f^s_{\bar L})$, $\; (f^d_{\bar R})$,
respectively), the flux density of axial number into the symmetric phase in
the plasma frame will be described by the expression (compare with
\cite{CKN92,AB94}):

\begin{eqnarray}
f_A = f_L^d + f_{\bar{L}}^s - f_R^s -f_{\bar{R}}^d =
 \nonumber \\
= \frac{2}{\gamma}\int\limits_0^\infty dk_l \int\limits_0^\infty
\frac{k_tdk_t}{4\pi^2}
\frac{k_l}{k} \cdot
\left\{ e^{ \gamma \cdot \left( E+v_wk_l \right) /T } + 1 \right\} ^{-1}
\cdot \left[ K_+-K_- \right] ,
\label{4.1}
\end{eqnarray}
where $v_w$ is the wall velocity, $\gamma$ is the Lorentz-factor and

\begin{equation}
K_+ - K_- = h''K =
\frac{\eta ''(1+\eta^2)}{\eta (1+\eta ''^2)}\cdot\frac{(G^2 - F^2)}{2}.
\label{3.22}
\end{equation}
(see also (\ref{3.3}), (\ref{3.12}), (\ref{3.13})), while all quantities in
these formulae can be expressed via $k_l$ and $k_t$.  Since the isosinglets
and isodoublets posses the different values of weak hypercharge $(Y_d = 4/3,
\; Y_s = 1/3)$, which is conserved in symmetric phase, the flux density of
weak fermionic hypercharge emitted from the bubble into the old phase will
be

\begin{equation}
f_Y = \frac{4}{3}\left( f^d_L - f^d_{\bar{R}} \right) +
\frac{1}{3}\left( f^s_R - f^s_{\bar{L}} \right) = \frac{1}{2}f_A.
\label{4.2}
\end{equation}

As it was shown in \cite{CKN92,AB94}, a nonzero hypercharge biases an
anomalous weak interactions towards production of baryons rather of
antibaryons. This processes convert the ($P$ and $CP$)-asymmetric state into
($C$ and $CP$)-asymmetric one. The net baryon number produced in a unit
volume

\begin{equation}
n_B \sim \frac{6\Gamma_B}{T^3}\cdot\frac{f_Y\tau_T}{v_w} ,
\label{4.3}
\end{equation}
where

$$
\Gamma_B \sim a(\alpha_WT)^4
$$
is the rate per unit volume of baryon violating transitions,
factor $a\sim (0.1 \div 1.0), \; \alpha_w^4 \sim 10^{-6}, \; \tau_T$ ---
"transport time", i.e. the average time each transmitted fermion spends in
the old phase before being struck by the bubble wall for a second time.
Using the expression for entropy

$$
s = 2\pi^2g_\ast\frac{T^3}{45},
$$
where $g_\ast \sim 100$ is a statistical factor, for the baryon-to-entropy
ratio produced during the phase transition finally we obtain:

\begin{equation}
\delta = \frac{n_B}{s} \sim \frac{6a\alpha_W^4Tf_Y\tau_T}{sv_w} \sim
\frac{150a\alpha_W^4}{\pi^2g_\ast}\cdot\frac{f_Y\tau_T}{v_wT^2} .
\label{4.4}
\end{equation}
Let us evaluate this quantity. If we consider a $top$-quark as a fermion,
responsible for BAU generation, then numerical estimations show, that
$\tau_T \sim O(100/T)$ \cite{CKN93}. The bubble walls with a width smaller
than a particle mean free path move with a velocity $v_w \sim 0.04$
\cite{HKLLM92}.  The phase transition takes place at temperature $T\sim 10^2
GeV$. Thus we can write that

\begin{equation}
\delta \sim (10^{10} \div 10^{-11})(GeV)^{-3}\cdot f_A .
\label{4.5}
\end{equation}
Numerical estimations of the integral (\ref{4.1}) give

\begin{equation}
f_A \sim  (10^3 \div 10^4)  (GeV)^3,
\label{4.6}
\end{equation}
As it is easy to see, this value does correspond to the observed value
(\ref{1.1}).

Thus we can conclude, that the rate of $CP$-violation through the mechanism
described above is sufficient enough to explain the BAU. Since all
calculations were produced in the frames of MSM, it means that there is no
reason to take different extensions of this model in search of extra source
for $CP$-violation. However still remains the problem of too small Higgs
mass ($m_H < 45GeV$) necessary for the produced BAU not to be eliminated by
the $B$-violating processes in the broken phase. Thus the question of
validity of MSM for baryogenesis remains open.

Modifications of the $CP$-violation mechanism described above can be applied
also to the models of "spontaneous baryogenesis" \cite{CKN93},
of baryogenesis from cosmic strings \cite{B} and other models. We are going
to investigate them in future papers.

%%%%%%%%%%%%%%%%%%%%%%%%%%%%%%%%%%%%%%%%%%%%%%%%%%%%%%%%%%%%%
\vskip 0.5cm
{\Large \bf Acknowledgments.}
\vskip 0.3cm

Authors would like to thank Yu.Werbetsky, O.Kancheli, I.Gogoladze and
A.Kobakhidze for useful discussions and N.Shubitidze for his help in
numerical calculations.

The research described in this publication was made possible in part by
Grant MXL000 from the International Science Foundation.

%%%%%%%%%%%%%%%%%%%%%%%%%%%%%%%%%%%%%%%%%%%%%%%%%%%%%%%%%%%%%%%%%%%%

\end{document}